\documentclass[%
 reprint,
 amsmath,amssymb,
 aps,
]{revtex4-2}

\usepackage{lineno}
\usepackage{graphicx}
\usepackage{dcolumn}
\usepackage{bm}
\usepackage{float}

\begin{document}

\preprint{APS/123-QED}

\title{From Turbulence to Landscapes: Universality of Logarithmic Mean Profiles in Bounded Complex Systems}

\author{Milad Hooshyar}
 \affiliation{Princeton Environmental Institute and Princeton Institute for International and Regional Studies, Princeton University, USA}
 
 \author{Sara Bonetti}
 \affiliation{Department of Environmental Systems Science, ETH Zurich, Switzerland}
  
 \author{Arvind Singh}
 \affiliation{Department of Civil, Environmental, and Construction Engineering, University of Central Florida, USA}

\author{Efi Foufoula-Georgiou}
\affiliation{Department of Civil and Environmental Engineering and Department of Earth System Science, University of California, Irvine, USA}
 
 \author{Amilcare Porporato}
 \email{Corresponding author, aporpora@princeton.edu}
 \affiliation{Princeton Environmental Institute and Department of Civil and Environmental Engineering, Princeton University, USA}
 
\date{\today}

\begin{abstract}
The logarithmic mean-velocity profile is a key experimental and theoretical result in wall-bounded turbulence. Similarly, here we show that the topographic surface emerging between parallel zero-elevation boundaries presents an intermediate region with a logarithmic mean-elevation profile. We use model simulations, which account for growth, erosion, and smoothing processes and give rise to complex topography with channel branching and fractal river networks, as well as data from a physical landscape-evolution experiment. Dimensional and self-similarity arguments are used to corroborate this finding. Our results suggest a universality of the  logarithmic scaling in bounded complex systems out of equilibrium, of which landscape topography and turbulence are quintessential examples. 
\end{abstract}

\maketitle


The striking channel and ridge patterns in the land surface emerge from a competition between growth and erosion, driven by external climatic and tectonic forcings, resulting in surfaces that exhibit several well-known scaling laws \citep{Rodriguez2001}. The complex networks of channels and their self-similar statistical properties have common features with other branch-forming systems \citep{kramer1992evolution, arneodo1992golden} and have become a key example of out-of-equilibrium systems in statistical physics \citep{sinclair1996mechanism, rinaldo1996thermodynamics, banavar1997sculpting, banavar2001scaling}. The resemblance between the landscape self-similarity and the scale invariance observed in turbulence has been exploited to analyze landscape morphology \citep{paola1996incoherent, passalacqua2006application}. In particular,  \citet{bonetti2019Cascade} emphasized the parallels between the channelization cascade and the hierarchical pattern formation toward finer scales observed in the non-equilibrium systems such as hydrodynamic turbulence \citep{tennekes1972first, chavarria1995hierarchy, frishman2018turbulence}. 

Here this analogy between turbulence and landscape evolution is strengthened by the discovery of a logarithmic region in the mean-elevation profiles, which resembles the logarithmic scaling of the mean stream-wise velocity in the intermediate region of wall bounded turbulent flows \citep{tennekes1972first,bradshaw1995law, jimenez2012cascades, luchini2017universality}. 
As well known, in wall-bounded turbulence the mixing-length argument by \citet{prandtl19257} first justified a logarithmic velocity profile which was then generalized by means of dimensional analysis and similarity considerations \citep{millikan1938,barenblatt1996scaling, luchini2017universality}. The constancy of the slope of the logarithmic profile, commonly denoted by $1/\kappa$, where $\kappa$ is the von Karman's constant, has been verified in numerous experimental and numerical studies \citep{Nikuradse, eggels_1994, bailey_vallikivi_hultmark_smits_2014, luchini2017universality}. 

We establish the existence of a logarithmic region in mean-elevation profiles using a combination of numerical simulations, laboratory experiments, and dimensional considerations. For high erosion rates, we obtain  mean-elevation profiles consisting of a linear part close to the fixed-elevation boundary, similar to the viscous sub-layer in turbulence \citep{mckeon2004further},  followed by a logarithmic profile in the intermediate  distant from both the boundary and the central part of the domain, similar to the inertial sub-layer in turbulence \citep{tennekes1972first, marusic2013logarithmic}. 

{\it Governing equations. --} The evolution of the land surface elevation in detachment-limited condition \citep{howard1994detachment, izumi1995inception, perron2008controls} results from a balance between diffusive soil creep, fluvial erosion, and uplift,  

\begin{equation}
\label{eq:LEM_1}
\frac{\partial z}{\partial t}=D\Delta z-K a^m |\nabla z|^n + U,
\end{equation}
where $z\left(x, y, t\right)$ is the surface elevation at time $t$ and location $(x, y)$, $D$ is soil diffusivity, and $U$ is the tectonic uplift rate. The term $K a^m |\nabla z|^n$ is a sink term, which quantifies fluvial erosion with parameters $K$, $m$, and $n$. The variable $a(x, y, t)$ is defined as the contributing area per unit contour-line length. The presence of $a$ in Eq. \eqref{eq:LEM_1} follows from a condition of steady-state water depth generated by a unit rainfall rate under the assumption that water moves in the direction of the local slope with a constant velocity. As a result \citep{bonetti2018theory,bonetti2019Cascade}, $a$ is given by the equation

\begin{equation}
\label{eq:water_depth_1}
\nabla \cdot\left(a\frac{\nabla z}{|\nabla z|}\right) = -1.
\end{equation}

By non-dimensionalizing the system of equations \eqref{eq:LEM_1} and \eqref{eq:water_depth_1}, \cite{bonetti2019Cascade} derived the dimensionless number $\chi$ which quantifies the relative impact of fluvial erosion to diffusive transport and uplift, 

\begin{equation}
\label{eq:chi}
\chi =  \frac{K l^{m+n}}{D^n U^{1-n}}, 
\end{equation}
where $l$ is a typical length scale of the domain. In this study we focus on the case $n=1$; different values of $n$ do not qualitatively change our results, as discussed in the Supplemental Material.  

\begin{figure*}[t]
  \centering
  \includegraphics[width=0.8\textwidth]{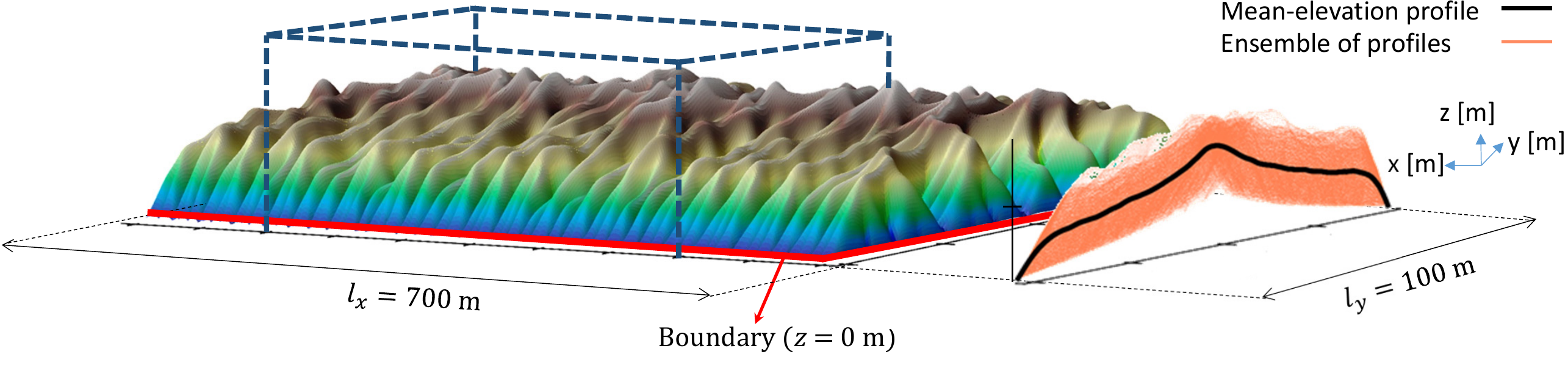}
  \caption{An example of the steady-state surface from the numerical solution of equations \eqref{eq:LEM_1} and \eqref{eq:water_depth_1}. The simulation domain is $l_x = 700\ \text{m}$ by $l_y=100\ \text{m}$ and has $1\ \text{m}$ grid spacing. The boundary conditions at the edges are constant elevation $z = 0\ \text{m}$. On the $y-z$ plane, the mean-elevation profile and the ensemble of profiles are shown. To minimize the effect of domain size along x-axis, for further analysis we used the surface within $100\ \text{m}\leq x \leq 600\ \text{m}$ as highlighted by the box.}
  \label{fig:sim_domain}
\end{figure*}

In analogy with a channel flow between parallel plates, we numerically solved the system of equations \eqref{eq:LEM_1} and \eqref{eq:water_depth_1} in a strip of land, unbounded in the $x$ direction and with constant elevation at two sides, $z(x, 0)=z(x, l_y)=0\ \text{m}$, where $l_y$ is the width in the direction $y$ (see Fig. \ref{fig:sim_domain}a and the Supplemental Material for more details on numerical simulations).  After reaching steady state, the mean-elevation profile $\Bar{z}(y)$  was obtained by averaging the $z$ field along the $x-$axis for $100\ \text{m} \le x \le 600\ \text{m}$ (Fig. \ref{fig:sim_domain}) to minimize the effect of  $l_x$  on the mean behavior of the surface. An example of mean-elevation profile (solid black line) and the ensemble of profiles (red lines) are shown in $y-z$ plane of Fig. \ref{fig:sim_domain}. \\

\begin{figure*}[t]
  \centering
  \includegraphics[width=0.8\textwidth]{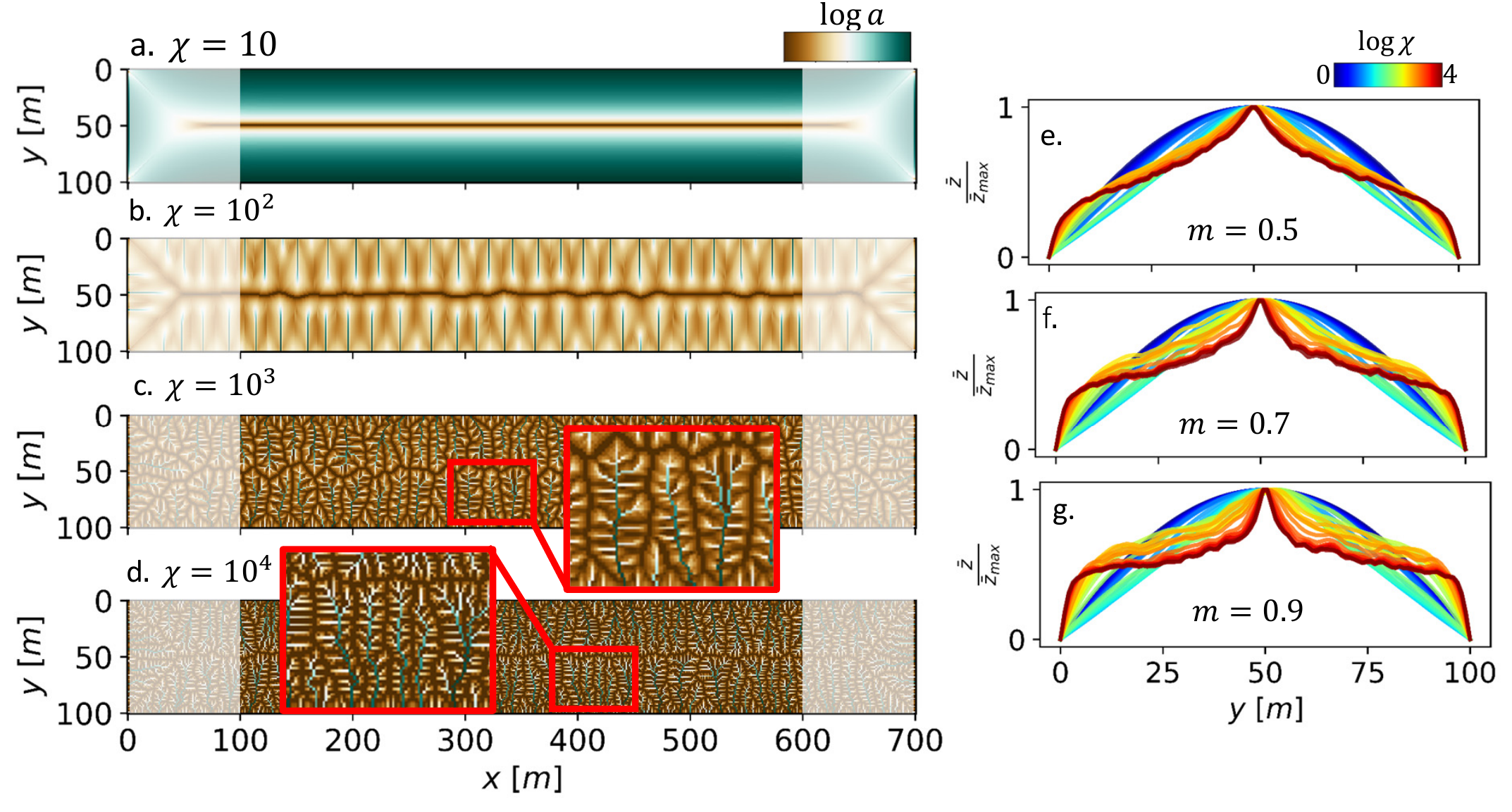}
  \caption{The landscape surface and mean-elevation profile for different $\chi$ and $m$. (a-d) The specific catchment area $a$ for $\chi=10$, $10^2$, $10^3$, and $10^4$ and $m = 0.5$. The shaded segments at the boundaries are discarded for further extraction of mean-elevation profiles. The red boxes in c and d show zoomed-in channel patterns in the marked portions of the domain. (e-g) The average profile for a wide range of $\chi$ and $m = 0.5$, $m=0.7$, and $m = 0.9$. As $\chi$ increases (higher relative proportion of fluvial erosion to diffusive transport), the surface becomes more dissected with branching channels. This results in the evolution of the mean-elevation profile toward a flatter shape.}
  \label{fig:profile_X}
\end{figure*}

Increasing $\chi$ (i.e., increasing relative magnitude of fluvial erosion) results in a more dissected surface with branching channels (see \citet{bonetti2019Cascade}). As shown in Fig. \ref{fig:profile_X}, for $\chi = 10$, the emerged surface is smooth with no channels as the diffusive transport is dominant and prevents the growth of instability and channel formation. In these conditions, the mean elevation profile can be obtained analytically for $m=n=1$ in terms of a hypergeometric function \citep{bonetti2019Cascade}. As $\chi$ exceeds a critical value ($\approx 37$ from linear stability analysis for $m=n=1$), parallel channels emerge and, with higher $\chi$, the surface becomes further dissected with the development of secondary branches \citep{bonetti2019Cascade}. The formation of branching channels impacts the mean-elevation profile as shown in Fig. \ref{fig:profile_X}$e-g$. It is evident that as $\chi$ increases the profiles become more uniform, similar to the flattening of the mean-velocity profile with increasing Reynolds number in turbulence \citep{Kundu2011}. 

{\it Dimensional Analysis and Self-Similarity. --} When averaged along the $x$ direction, the surface properties (e.g., elevation and slope) only depend on $y$. Therefore, the elevation field $z\left(x, y\right)$ can be decomposed into the sum of the mean elevation $\Bar{z}$ and fluctuations $z'$ around the mean, similarly to the Reynolds decomposition \citep{reynolds1895iv, paola1996incoherent}, 

\begin{equation}
\label{eq:decomp}
z\left(x, y\right) = \Bar{z}\left(y\right) + z', 
\end{equation}
where $\Bar{z}\left(y\right) = \lim_{l_x \to\infty}\frac{1}{l_x}\int_{0}^{l_x}z\left(x, y\right) dx$. The mean slope of steady-state surface is controlled by the parameters which describe the diffusive transport, fluvial erosion, and tectonic activity  ($D$, $K$, $U$, and $m$  in Eq. \eqref{eq:LEM_1} with $n=1$), the distance $y$ from the boundary, and two length scales in the $y$ and $z$ directions  ($l_y$ and $z_*$) \citep{barenblatt1996scaling},

\begin{equation}
 \label{eq:dz_dy_1}
 \frac{d \Bar{z}}{d y} = f_1\left(y, l_y, z_*, D, K, U, m\right).
\end{equation}

Choosing $y$, $D$, and $z_*$ as fundamental, dimensionally independent variables, the Pi-theorem yields

\begin{equation}
\label{eq:dimention_1}
\frac{y}{z_*}\frac{d \Bar{z}}{d y} = f_2\left(\frac{y}{l_y}, \frac{K y^{m+1}}{D}, \frac{U y^2}{D z_*}, m\right),
\end{equation}
from which simple manipulation of the variables leads to

\begin{equation}
\label{eq:dimention_2}
\left(m + 1\right) \eta\frac{d \phi}{d \eta} = f_3\left(\eta, \chi,\zeta, m\right), 
\end{equation}
where $\chi$ is given in Eq. \eqref{eq:chi} with $n=1$ and $l=l_y$ and $\phi=\frac{\Bar z}{z_*}$ is the normalized mean elevation $\Bar z$ by a factor $z_*$ which describes the overall elevation of the profile. Here, we used $z_*=\Bar z_{max}$, where $\Bar z_{max}$ is the average elevation at the divide ( $y=l_y/2$). The dimensionless value  $\chi$ is a global quantity (independent of $y$) reflecting the relative impact of fluvial erosion to diffusive transport \citep{bonetti2019Cascade}. The quantity $\eta=\frac{K y^{m+1}}{D}$ is a local variable with a similar form as $\chi$ but capturing the local relative contribution of those two processes, while $\zeta=\frac{U l_y^2}{D z_*}$ describes the relative impact of tectonic uplift to diffusive transport.  

In a system with relatively small diffusive transport and dominated by fluvial erosion and uplift, $\chi$ and $\zeta$ take high values. The same argument also applies to $\eta$ except for locations close to the boundary. Thus, when the variables $\eta$, $\chi$, and $\zeta$ reach such an asymptotic condition one may assume complete self-similarity \citep{barenblatt1996scaling} according to which the function $f_3$ is independent of these quantities

\begin{equation}
\label{eq: log_1}
\eta\frac{d \phi}{d \eta} = \kappa\left(m\right),
\end{equation}
where $\kappa$ is only a function of $m$. Integrating Eq. \eqref{eq: log_1} yields

\begin{equation}
 \label{eq:log_2}
 \phi =\kappa \left(m\right) \ln{\eta} + C, 
\end{equation}
where $C$ is independent of $\eta$ but may still depend on $m$, $\chi$, and $\zeta$. Eq. \eqref{eq:log_2} describes the logarithmic scaling of the mean-elevation profile with respect to $\eta$. The emergence of such a logarithmic profile is expected in systems dominated by fluvial erosion and away enough from both the boundary and the center of symmetry (high $\chi$, $\zeta$, and $\eta$).  

\begin{figure*}[t]
  \centering
  \includegraphics[width=0.8\textwidth]{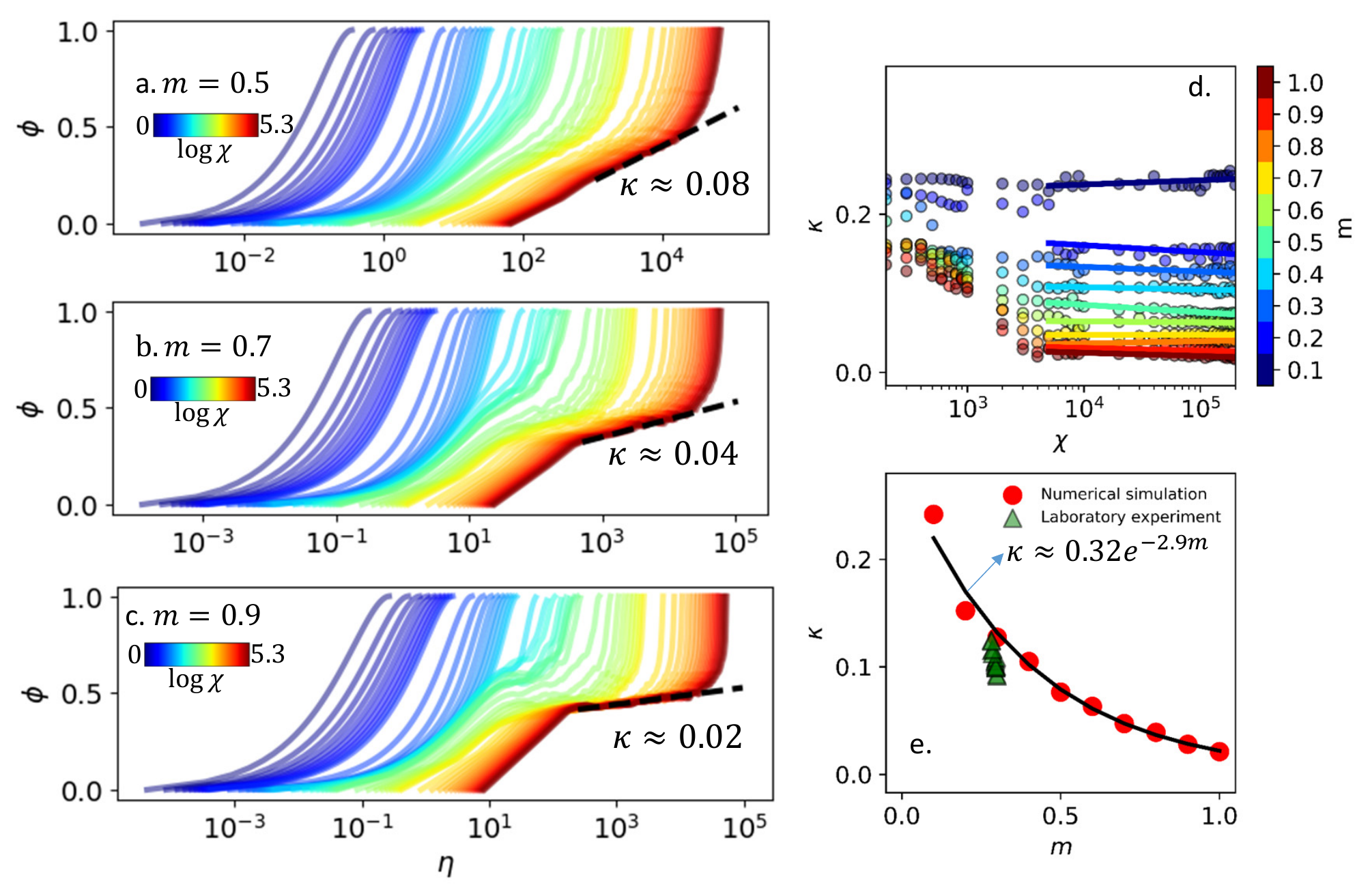}
  \caption{The logarithmic profile in numerical simulations. (a-c) The dimensionless profiles corresponding to $0\leq y \leq l_y/2$ in a semi-log space for $m = 0.5$, $0.7$, and $m = 0.9$. The dimensionless quantities are defined as  $\eta=\frac{K y^{m+1}}{D}$ and $\phi=\frac{z}{z_*}$ with $z_*=\Bar z_{max}$. Increasing $\chi$ leads to the emergence and further expansion of the logarithmic segment in the profiles. The dashed lines are the fitted logarithmic lines to the profiles with $\chi = 2 \times 10^5$ and their slopes are reported as $\kappa$. (d) The slope $\kappa$ of the logarithmic segment versus $\chi$ for different $m$. The slope $\kappa$ shows an asymptotic behavior with respect to $\chi$ and is independent of $\chi$ for relatively high $\chi$. The lines show the linear fits to the $(\chi, \kappa)$ data points for $\chi \geq 10^4$ at a given $m$ and their negligible slope confirms that  $\kappa$ is independent of $\chi$ for relatively high $\chi$. (e) The relationship between $\kappa$ and $m$. The slope $\kappa$ (average of simulations with $\chi \geq 10^4$) decreases monotonically with $m$ meaning that higher values of $m$ result in flatter logarithmic profile. The data points from a physical experiment (refer to the Supplemental Material for details) are also shown and lay close to the results from the numerical simulation.}
  \label{fig:dimensionless_profile}
\end{figure*}

{\it Logarithmic profile from simulations. --}  Fig. \ref{fig:dimensionless_profile}a-c shows the mean-elevation profile for a range of $\chi$ using dimensionless variables $\phi$ and $\eta$ with $z_*=z_{max}$ in a semi-log space for $m = 0.5$, $0.7$, and $0.9$. Given the symmetric simulation domain, here we show half of the profile  corresponding to $0\leq y \leq \frac{l_y}{2}$ , i.e. from boundary to the main drainage divide. The tendency of the mean-elevation profile toward a flatter shape and the logarithmic scaling (linear segment in the semi-log space of  Fig. \ref{fig:dimensionless_profile}a-c) clearly appear beyond a certain $\chi$. The logarithmic fits to the intermediate segments of the profiles for $\chi = 2 \times 10^5$ are shown in Fig. \ref{fig:dimensionless_profile}a-c (refer to the Supplemental Material for details). For a given $m$ and as $\chi$ increases, the logarithmic segments expand and their slope decreases with an asymptotic behavior in which $\kappa$ approaches to a slope independent of $\chi$ for relatively high $\chi$ (see Fig. \ref{fig:dimensionless_profile}d). 

{\it Laboratory Experiments. --} We also analyzed the data from a landscape evolution experiment performed using the eXperimental Landscape Evolution (XLE) facility in the St.  Anthony Falls Laboratory at the University of Minnesota, described in details in the Supplemental Material \citep{singh2015landscape, hooshyarclimatic}. The parameters modulating fluvial erosion and diffusion needed to analyze the logarithmic scaling were estimated assuming a steady state governed by equations \eqref{eq:LEM_1} and \eqref{eq:water_depth_1} with $n=1$ as explained in the Supplemental Material.  Having $D$, $K$, and $m$ from the experimental surfaces, we computed the slope of the logarithmic profile $\kappa$ for the 16 surfaces with the same optimization algorithm used to analyze the surfaces from numerical simulation. The surfaces from the physical experiment also contain the logarithmic scaling (see Fig. S1d and e) with slopes $\kappa$ that lay close to the results from the numerical simulation (Fig. \ref{fig:dimensionless_profile}d).

{\it Conclusions. --} 
Inspired by the resemblance between the progressive surface dissection and flattening of mean-elevation profile with the turbulence cascade and the turbulence velocity profile \citep{reynolds1895iv, Kundu2011}, we explored the existence of a logarithmic region in mean-elevation profile similar to the logarithmic scaling of stream-wise velocity  in wall bounded  turbulent flows \citep{tennekes1972first, banerjee2013logarithmic, marusic2013logarithmic, luchini2017universality}.

In turbulence, the logarithmic scaling in the inertial sub-layer is a fundamental result \citep{bradshaw1995law} associated to several other generic properties such as the nearly constant normalized Reynolds shear stresses \citep{pope2001turbulent}, the balance of the production and dissipation of turbulent kinetic energy \citep{pope2001turbulent, banerjee2013logarithmic}, and a characteristic behavior in terms of turbulence energy cascade within the inertial sub-range \citep{tennekes1972first}. Such a logarithmic scaling has also been reproduced under statistical stability through maximizing viscous dissipation \citep{malkus1989upper, bertram2015maximum}.

In landscape evolution, the existence of similar logarithmic scaling  may also be justified by invoking a competition between surface forming mechanisms and the tendency toward an optimal state \citep{Rigon1993, rinaldo1996thermodynamics, sinclair1996mechanism, Rodriguez2001}. The fact that we found a logarithmic region in both a minimalist model of landscape evolution and in laboratory experiments hints at the generality of such a scaling. A similar robustness to both boundary conditions and physical processes is present in turbulent velocity profiles, where a logarithmic region appears in both smooth and rough walls, as well as within different levels of  approximation (e.g., direct numerical simulations and large-eddy simulation) \citep{kim1987turbulence, jimenez2012cascades, cheng2014power}. Finally, our results from numerical simulations show that the logarithmic scaling persists for a wide range of model parameters, in agreement with the existence of the logarithmic scaling in turbulent flow for different types of fluids (e.g., Newtonian and non-Newtonian) \citep{wilson1985new, rudman2006direct}. 
In both turbulence and landscapes the logarithmic scaling emerges at an intermediate distance, when transitioning from an outer region toward the boundary. Landscapes and turbulence are quintessential examples of complex systems out-of-equilibrium, both exhibiting a cascade of patterns toward finer scales \citep{reynolds1895iv, bonetti2019Cascade}. Our results suggest a universality of the logarithmic scaling in such type of systems, where the progression of complexities (patterns) is prevented by the system boundary.

{\it Acknowledgements. --}  A.P. acknowledges support from the US National Science Foundation (NSF) grants EAR-1331846 and EAR-1338694, and BP through the Carbon Mitigation Initiative (CMI) at Princeton University. M.H acknowledges support from the Princeton Institute for International and Regional Studies (PIIRS) and the Princeton Environmental Institute (PEI). E.F. acknowledges support from the US National Science Foundation (NSF) grants DMS-1839336, EAR-1242458, and EAR-1811909.

\section{Supplemental Material} 

Here, we provide more details on the numerical and laboratory experiments used to establish the existence of the logarithmic scaling in mean-elevation profile. We also show the robustness of such scaling with respect to the parameter $n$ (the exponent of slope in the the governing equation).  

\subsection{Numerical Simulation}

We solved numerically the equations (1) and (2) to examine the mean-elevation profile under different conditions of diffusive transport and fluvial erosion. The simulation domain was a $700\ \text{m}$ by $100\ \text{m}$ rectangular grid with constant elevation $z=0\ \text{m}$ at the boundary, as shown in Fig. 1. This choice allowed us to neglect the effect of the domain dimension along the $x$ axis, mimicking the case of an infinite strip. 

We used the $D_{\infty}$ flow-direction algorithm to efficiently calculate the contributing area at each pixel in the discretized domain and then divided it by grid size to compute the specific catchment area  $a$ \citep{tarboton1997new, bonetti2018theory, bonetti2019Cascade}. The spatial grid spacing was $1\ \text{m}$ and the timestep of discretization $\Delta t$ was selected small enough to avoid numerical instabilities. As initial condition, we used a tent-shaped  surface plus random noise in which the local minima were filled. The steady-state condition was assumed to be reached when the average elevation change was less than $0.0001$ of the change due to the uplift ($U \Delta t$). For the diffusion term we used an explicit central finite difference approximation in space and a forward approximation in time. We used a semi-implicit scheme for time integration of the sink term by iteratively processing computational nodes starting from the nodes at the boundary, with $z=0\ \text{m}$ and no other node receiving their flow, and moving up toward the source nodes \citep{braun2013very}. 
We performed simulations with  $1\leq \chi \leq 2 \times 10^5$,  $0.1\leq m \leq 1$, and $0.8\leq n \leq 1.2$ to cover the range of parameters reported in the literature \citep{perron2008controls, sweeney2015experimental}. For simplicity, we used $D=0.005\ \mathrm{m}^2/\mathrm{yr}$ and changed $K$ to achieve desired $\chi$ for a given $m$ value according to Eq. (3). 

After reaching steady-state, we computed the characteristics of the logarithmic segment of the non-dimensional mean-elevation profile by fitting a function to the profile within $0\leq y \leq l_y/2$. This includes a linear part for the segment close to the boundary and a logarithmic function for the intermediate segment. We also considered an additional power function for the segment close to the divide to cover the whole profile and close the fitting problem. We found the best fits by maximizing the summation of $R^2$ values of the fit to each segment. The variables for optimization are two thresholds on $\eta$ which correspond to transition from the linear segment close to the boundary to the intermediate logarithmic segment, and the transition from logarithmic segment to the section close to the divide. Using such thresholds, simple linear regression gives the slope of the logarithmic profile denoted by $\kappa$ (see Eq. (9)).

\subsection{Laboratory Experiments}\label{exp_section}

\begin{figure*}[t]
  \centering
  \includegraphics[width=0.7\textwidth]{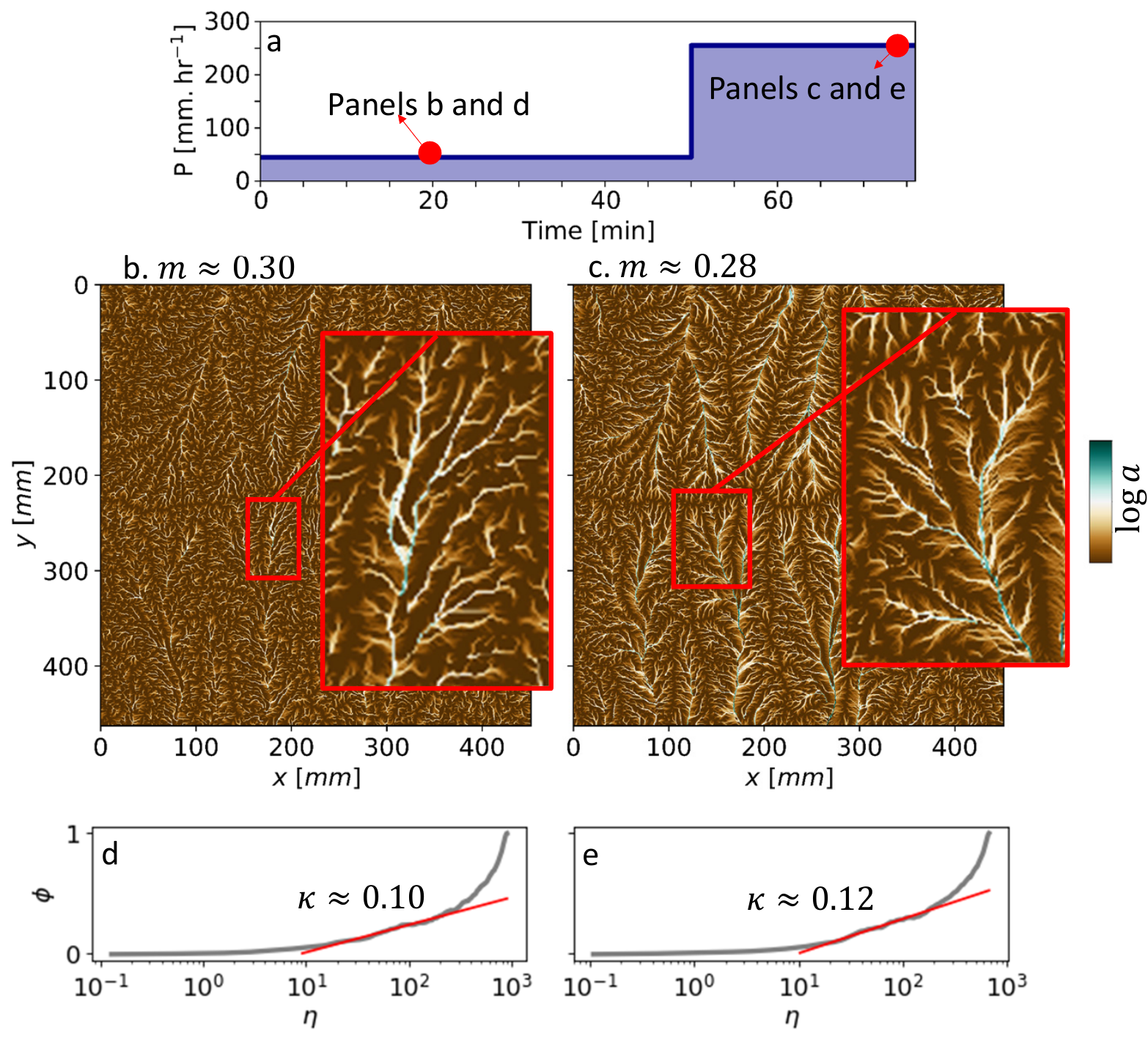}
  \caption{The logarithmic profile from the experimental landscape. (a) The time variation of rainfall throughout the simulation. The experiment was started with  $P=45\ \text{mm/hr}$ which was increased to $P=225\ \text{mm/hr}$. (b and c) The specific catchment area $a$ for two surfaces marked in 
  (a). (d and e) The mean-elevation profile represented by the non-dimensional quantities $\eta$ and $\phi$. The fitted lines to the logarithmic segment of the profiles are shown by the red lines with the slope reported as $\kappa$.}
  \label{fig:experiment}
\end{figure*}

We also analyzed the data from a landscape evolution experiment performed using the eXperimental Landscape Evolution (XLE) facility in the St.  Anthony Falls Laboratory at the University of Minnesota \citep{singh2015landscape, hooshyarclimatic}.  The experiment consists of a $500\ \text{mm}$ long, $500\ \text{mm}$ wide, and $300\ \text{mm}$ deep erosion box filled with a homogeneous mixture of fine silica and a surmounted rainfall simulator. The domain was subject to constant uplift rate ($U = 20\ \text{mm/hr}$) and initial constant rainfall rate of $P=45\ \text{mm/hr}$ which was increased to $P=225\ \text{mm/hr}$ (see Fig. \ref{fig:experiment}a). We analyzed 16 surfaces ($0.5 \ \text{mm}$ by $0.5\ \text{mm}$ Digital Elevation Models) scanned at 5 min intervals spanning 75 min of the simulation. The first 10 surfaces correspond to $P=45\ \text{mm/hr}$ and the last 6 to $P=225\ \text{mm/hr}$. Fig. \ref{fig:experiment} shows two examples of the experimental surface. 

The parameters modulating fluvial erosion and diffusion needed to analyze the logarithmic scaling were estimated assuming a steady state governed by equations (1) and (2) with $n=1$ \citep{sweeney2015experimental}. The specific catchment area $a$ was computed by dividing the total contributing area from $D_\infty$ by the grid spacing of the scanned surfaces ($0.5\ \text{mm}$) \citep{tarboton1997new}. Following \citet{perron2009formation} and focusing on the regions with small fluvial erosion (hilltops with small $a$ and $|\nabla z|$), from Eq. (1), $D$ was approximated as

\begin{equation}
\label{eq:D_approx}
D= -\frac{U}{{\triangle z}_{a|\nabla z|\xrightarrow{}0}}.
\end{equation}

Given $D$ in Eq. (1) at steady state, $K$ and $m$ can be estimated by fitting a  power function to $\frac{D\triangle z + U}{|\nabla z|}$ versus $a$ relationship \citep{perron2009formation, hooshyarclimatic}

\begin{equation}
\label{eq:Km_approx}
\frac{D\triangle z + U}{|\nabla z|}= K a^{m}.
\end{equation}

Having $D$, $K$, and $m$ from the experimental surfaces, we computed the slope of the logarithmic profile $\kappa$ for the 16 surfaces with the same optimization algorithm used to analyze the surfaces from numerical simulation. Examples of the fitted logarithmic profiles are shown in Fig. \ref{fig:experiment}d and e. \\

\subsection{The effect of $n$ on the logarithmic scaling}
\begin{figure*}[t]
  \centering
  \includegraphics[width=0.7\textwidth]{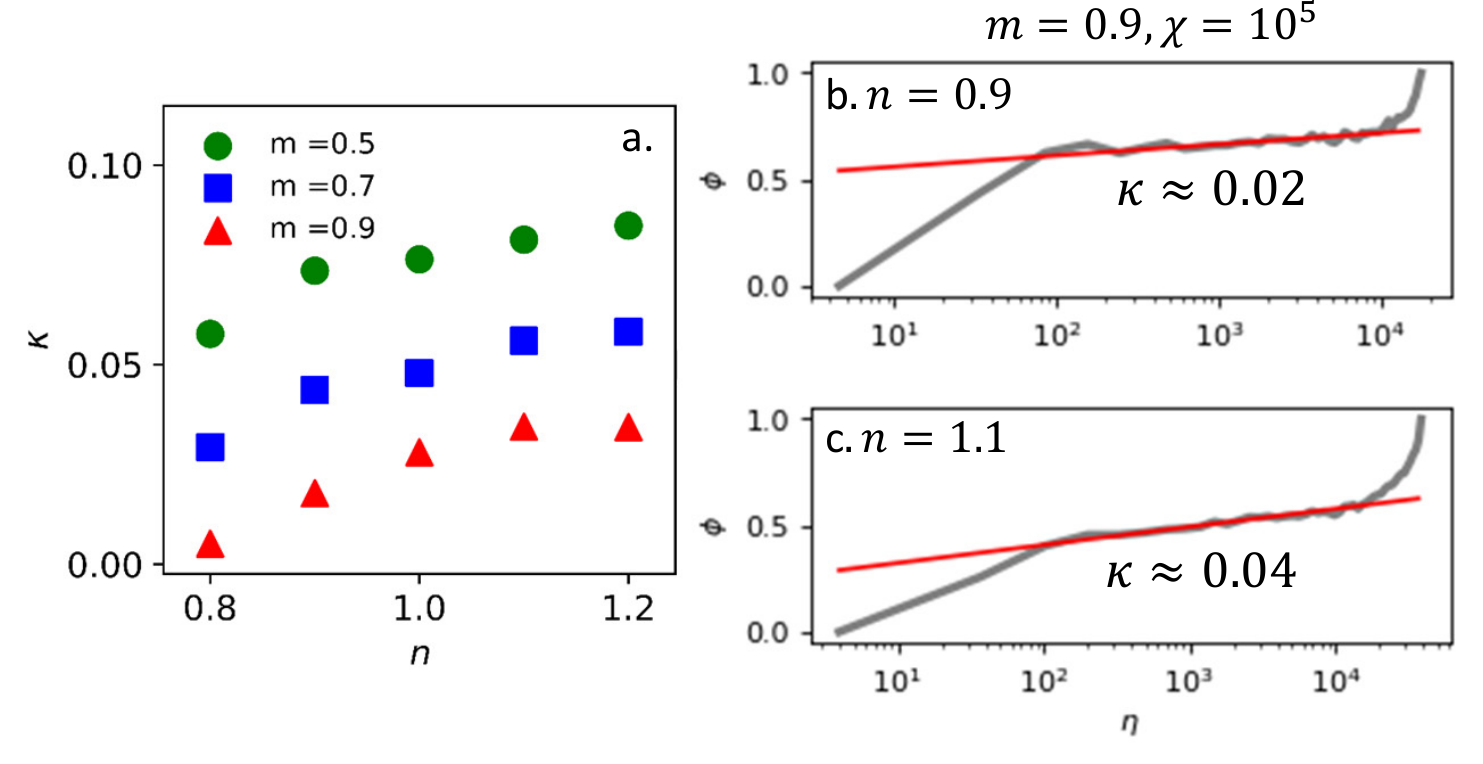}
  \caption{The dependence of the slope of logarithmic scaling on the parameter $n$. (a) The logarithmic scaling persists for a range of $n$ and the slope $\kappa$ increases with higher $n$. (b and c) The non-dimensionalized mean-elevation profiles for two cases with the same $m=0.9$ and $\chi = 10^5$, but different $n$. In both cases the logarithmic scaling is easily detectable. }
  \label{fig:SI_effect_n}
\end{figure*}

Here, we investigate the controls of the parameter $n$ in Eq. (1) on the characteristics of the logarithmic scaling of mean-elevation profile. The self-similarity arguments presented in the paper for $n=1$ can be readily generalized with respect to $n$ by defining $\eta=\frac{K y^{m+n}}{D^nU^{1-n}}$ and using $\chi$ from Eq. (3) with $l=l_y$. Following the same line of reasoning, one can find

\begin{equation}
 \label{eq:SI_log}
 \phi =\kappa \left(m, n\right) \ln{\eta} + C, 
\end{equation}

where $\kappa$ is a function of both $m$ and $n$. To reveal  details on the functional dependence of $\kappa$ on $n$, we run an additional 300 simulations with $\chi \geq 10^4$, $m=0.5$, $0.7$, and $0.9$, $n=0.8$, $0.9$, $1$, $1.1$, and $1.2$, and the same boundary and initial condition discussed earlier. Our result clearly shows that , the logarithmic scaling emerges for a range of parameter $n$ (Figure \ref{fig:SI_effect_n}b and c) and the slope $\kappa$ increases with higher $n$ values (Fig. \ref{fig:SI_effect_n}a). \\

\bibliography{main.bib}

\end{document}